# The Wiener-Filtered COBE DMR Data and Predictions for the Tenerife Experiment

Emory F. Bunn[1], Yehuda Hoffman[2], and Joseph Silk[1]

1. *Departments of Astronomy and Physics and Center for Particle Astrophysics, University of California, Berkeley, CA 94720*

2. *Racah Institute of Physics, Hebrew University, Jerusalem, Israel*

## ABSTRACT

We apply a Wiener filter to the two-year COBE DMR data. The resulting sky map has significantly reduced noise levels compared to the raw data: the most prominent hot and cold spots are significant at the 4-$\sigma$ level. Furthermore, the entire covariance matrix of the errors in the filtered sky map is known, and it is therefore possible to make constrained realizations of the microwave sky with the correct *a posteriori* probability distribution. The filtered DMR sky map is used to make predictions for the Tenerife experiment. Two prominent features are predicted in a region of the sky not yet analyzed by the Tenerife group. The presence of these features is a robust prediction of the standard cosmological paradigm; if these features are not observed, some of our fundamental assumptions must be incorrect.
Subject headings: cosmic microwave background — methods: statistical

# 1 Introduction

The discovery of anisotropy in the cosmic microwave background (CMB) by the COBE satellite (Smoot et al. 1992) marked a major advance in cosmology. CMB anisotropy in general and the COBE data in particular now provide some of the most powerful tests of cosmological theories, particularly theories concerning the formation of large-scale structure.

The COBE DMR data have been extensively analyzed, with most of the emphasis being placed on the task of estimating the power spectrum of the anisotropy (*e.g.*, Bennett et al. 1994; Bond 1995; Bunn, Scott, & White 1995; Bunn & Sugiyama 1995; Górski et al. 1994; White & Bunn 1995) although a few have focused on other problems, such as testing the hypothesis that the anisotropy obeys Gaussian statistics (*e.g.*, Hinshaw et al. 1994, Kogut et al. 1995). In addition, statistical comparisons have been made between the DMR sky maps and data from other experiments (Ganga et al. 1993, Watson & Gutiérrez de la Cruz 1993, Lineweaver et al. 1995). Less attention has been paid to studying the spatial properties of the sky maps themselves to determine, for example, which hot and cold spots on the maps are likely to be real, and which are dominated by noise. Such an analysis has the potential to be quite useful in comparing the DMR results to other experiments that probe comparable angular scales, such as the FIRS (Meyer et al. 1991) and Tenerife (Davies et al. 1992, Hancock et al. 1994, Watson et al. 1992) experiments. In this paper we will use Wiener filtering to study these questions, extending a similar study that we did for the first-year DMR maps (Bunn et al. 1994). In particular, we will use the Wiener-filtered two-year COBE DMR sky maps to make predictions for the Tenerife experiment, which probes smaller (although comparable) angular scales than COBE.

There are two major difficulties in studying the DMR maps. The signal-to-noise ratio in each pixel is quite low, and pixels near the Galactic plane are contaminated by Galactic emission. These problems place strict limits on the accuracy with which the true structure can be recovered (Bunn, Hoffman, & Silk 1994; Stark 1993). The techniques of Wiener filtering and "constrained realizations" provide one way of mitigating these problems. We assume the correctness of the canonical theory of large-scale CMB anisotropy: that the anisotropy forms a Gaussian random field with a known power spectrum. We then remove the pixels that are presumed to be contaminated, and apply an optimal linear filter to the data in an attempt to clean up the noise and see the underlying structure. The technique of constrained realizations helps quantify the uncertainties associated with this method.

The Wiener filter described in this paper is based on techniques developed by Wiener (1949) and introduced to astrophysics by Rybicki & Press (1992), and it is similar to a technique applied recently to galaxy catalogues (Lahav et al. 1994, Zaroubi et al. 1994) The formalism for making constrained realizations of Gaussian random fields was developed by Hoffman & Ribak (1991) and has been applied to the study of large-scale structure by Ganon & Hoffman (1993).

The idea behind the Wiener filter is quite simple. If we are willing to assume that we know the statistical properties of the signal and noise contributions to the data, we can apply a linear filter that preferentially removes noise and leaves signal. The Wiener filter is the optimal linear filter for this purpose. It can be derived in two independent ways. First, it is the optimal linear filter in the sense of least squares. That is, the Wiener filter minimizes the expected mean-square deviation between the filtered data and the underlying cosmic signal. It is also a maximum-likelihood estimator of the cosmic signal. That is, the filtered sky map is the most likely realization of the underlying signal at each point. As long as both the signal and the noise are Gaussian, the filters chosen by these two criteria are identical.

Furthermore, we can determine the complete *a posteriori* probability distribution of the true signal given the data and the assumed power spectrum. This probability distribution is Gaussian, with mean given by the filtered data. We can compute the covariance matrix of this probability distribution and use it to construct "constrained realizations" of the CMB anisotropy with the correct *a posteriori* probability distribution. These realizations are useful for qualitatively assessing the uncertainties in the filtered maps, and for using Monte Carlo simulations to make predictions for other experiments.

Although detailed derivations of the Wiener filter have been given elsewhere (Bunn et al. 1994, Bunn 1995), we will give a brief sketch of the central results in the next section. We will then apply this formalism to the two-year DMR sky maps to identify features in the data that have high statistical significance. Finally, we will use the filtered data to make predictions for the Tenerife experiment.

## 2  Deriving the Wiener Filter

As usual, we will expand the temperature anisotropy $\Delta T$ in spherical harmonics:

$$\Delta T(\hat{\mathbf{r}}) = \sum_{l,m} a_{lm} Y_{lm}(\hat{\mathbf{r}}). \tag{1}$$

The signal seen by the COBE DMR experiment is the convolution of $\Delta T$ with a beam pattern $W$. The spherical harmonic coefficients of the DMR signal are therefore $W_l a_{lm}$, where $W_l$ is a coefficient of a Legendre polynomial expansion of the beam pattern. The DMR beam pattern is often approximated as a Gaussian with a $7°$ beam width, but we use the more accurate characterization given by Wright et al. (1994).

After removing all pixels within $20°$ of the Galactic plane, the COBE DMR data contain $\mathcal{N} = 4038$ pixels. Each pixel contains contributions from both signal and noise:

$$d_i = \sum_{l,m} W_l a_{lm} Y_{lm}(\hat{\mathbf{r}}_i) + n_i. \tag{2}$$

Here $d_i$ is the measured value of the $i$th pixel, $a_{lm}$ is a coefficient of a spherical harmonic expansion of the actual CMB anisotropy, $\hat{\mathbf{r}}_i$ is the location of the $i$th pixel on the sky, and $n_i$ is the noise. If we denote pairs of indices $(l, m)$ by a single Greek index $\mu$ and define matrices $\mathbf{Y}$ and $\mathbf{W}$ by $Y_{i\mu} = Y_\mu(\hat{\mathbf{r}}_i)$ and $W_{\mu\mu'} = W_l \delta_{\mu\mu'}$ [where $\mu = (l, m)$], then we can write the above equation in a more compact form:

$$\vec{d} = \mathbf{Y} \cdot \mathbf{W} \cdot \vec{a} + \vec{n}. \tag{3}$$

We make the usual assumption of Gaussian statistics: both $a_\mu$ and $n_i$ are Gaussian random variables with zero mean and covariances

$$\langle \vec{a} \cdot \vec{a}^T \rangle = \mathbf{C}, \tag{4}$$

$$\langle \vec{n} \cdot \vec{n}^T \rangle = \mathbf{N}, \tag{5}$$

$$\langle a_\mu n_i \rangle = 0. \tag{6}$$

The matrices $\mathbf{C}$ and $\mathbf{N}$ are diagonal. $C_{\mu\mu} = C_l$ is simply the angular power spectrum, which can be predicted from a particular theoretical model, and $N_{ii} = \sigma_i^2$ is the noise variance in the $i$th pixel. [The noise covariance matrix is not strictly diagonal, but the approximation that it is diagonal is quite good (Lineweaver et al. 1994).]

In principle, $\vec{a}$ and $\mathbf{C}$ should be infinite-dimensional, but in practice the COBE beam smooths away modes with high $l$, and we can truncate the spherical harmonic expansion at some $l_{\max}$. We shall see below that the filter recovers virtually no information for $l \gtrsim 20$, and so we generally set $l_{\max} = 30$. The dimension of $\vec{a}$ is therefore $\mathcal{M} = (l_{\max} + 1)^2 = 961$. The covariance matrix of the data vector $\vec{d}$ contains contributions from both signal and noise:

$$\mathbf{M} \equiv \langle \vec{d} \cdot \vec{d}^T \rangle = \mathbf{Y} \cdot \bar{\mathbf{C}} \cdot \mathbf{Y}^T + \mathbf{N}, \tag{7}$$

where the diagonal matrix $\bar{\mathbf{C}} = \mathbf{W} \cdot \mathbf{C} \cdot \mathbf{W}$ contains the beam-smoothed angular power spectrum.

The Wiener filter is the linear filter $\mathbf{F}$ that estimates the true cosmic signal $\vec{s} \equiv \mathbf{Y} \cdot \vec{a}$ as closely as possible, in the sense of least squares. Specifically, if we set

$$\vec{y} = \mathbf{F} \cdot \vec{d}, \tag{8}$$

then we want to choose the $\mathcal{N} \times \mathcal{N}$ matrix $\mathbf{F}$ to minimize

$$\Delta \equiv \langle (\vec{y} - \mathbf{Y} \cdot \vec{a})^2 \rangle. \tag{9}$$

Setting $\partial \Delta / \partial F_{ij} = 0$ and solving for $\mathbf{F}$, we find the least-squares filter

$$\mathbf{F} = \mathbf{Y} \cdot \mathbf{C} \cdot \mathbf{W} \cdot \mathbf{Y}^T \cdot \mathbf{M}^{-1}. \tag{10}$$

This derivation makes no use of the fact that the random variables $\vec{a}$ and $\vec{n}$ are Gaussian; it uses only their covariances. If we make use of our assumption that these quantities are Gaussian, we can give an alternative characterization of the filter (10): it gives the maximum-likelihood estimator of the signal $\vec{s}$.

Bayes's theorem states that the conditional probability density of $\vec{a}$ being the correct set of spherical harmonic coefficients given the data $\vec{d}$ is simply $f(\vec{a} \,|\, \vec{d}) \propto f(\vec{d} \,|\, \vec{a}) f(\vec{a})$, where $f(\vec{d} \,|\, \vec{a})$ is the probability density that the actual data $\vec{d}$ would have occurred assuming $\vec{a}$ is the correct set of coefficients, and $f(\vec{a})$ is the *a priori* probability of getting the coefficients $\vec{a}$. All of these probability densities are Gaussian:

$$f(\vec{a} \,|\, \vec{d}) \propto \exp\left(-\tfrac{1}{2}(\vec{d} - \mathbf{Y} \cdot \mathbf{W} \cdot \vec{a})^T \cdot \mathbf{N}^{-1} \cdot (\vec{d} - \mathbf{Y} \cdot \mathbf{W} \cdot \vec{a})\right) \exp\left(-\tfrac{1}{2} \vec{a}^T \cdot \mathbf{C}^{-1} \cdot \vec{a}\right). \tag{11}$$

By completing the square and performing some unenlightening algebraic manipulations, one can show that $f(\vec{a} \,|\, \vec{d}) \propto \exp\left(-\tfrac{1}{2}\chi^2\right)$, where

$$\chi^2 = (\vec{a} - \vec{a}_{\mathrm{ML}})^T \cdot (\mathbf{C}^{-1} + \mathbf{W} \cdot \mathbf{Y}^T \cdot \mathbf{N}^{-1} \cdot \mathbf{Y} \cdot \mathbf{W}) \cdot (\vec{a} - \vec{a}_{\mathrm{ML}}) + \vec{d}^T \cdot \mathbf{M}^{-1} \cdot \vec{d}, \tag{12}$$

and

$$\vec{a}_{\mathrm{ML}} = (\mathbf{C}^{-1} + \mathbf{W} \cdot \mathbf{Y}^T \cdot \mathbf{N}^{-1} \cdot \mathbf{Y} \cdot \mathbf{W})^{-1} \cdot \mathbf{W} \cdot \mathbf{Y}^T \cdot \mathbf{N}^{-1} \cdot \vec{d}. \tag{13}$$

It is easy to see from equation (12) that the maximum-likelihood value of $\vec{a}$ is $\vec{a}_{\mathrm{ML}}$. The maximum-likelihood estimate of the actual CMB signal is therefore $\vec{s}_{\mathrm{ML}} = \mathbf{Y} \cdot \vec{a}_{\mathrm{ML}} = \mathbf{F}' \cdot \vec{d}$, where the filtering matrix is

$$\mathbf{F}' = \mathbf{Y} \cdot \left(\mathbf{C}^{-1} + \mathbf{W} \cdot \mathbf{Y}^T \cdot \mathbf{N}^{-1} \cdot \mathbf{Y} \cdot \mathbf{W}\right)^{-1} \cdot \mathbf{W} \cdot \mathbf{Y}^T \cdot \mathbf{N}^{-1}. \tag{14}$$

The two filters $\mathbf{F}$ and $\mathbf{F}'$ are in fact identical, despite their different appearances. Four our purposes, since the number of data points $\mathcal{N}$ is larger than the number of degrees of freedom $\mathcal{M}$, equation (14) is more convenient computationally, requiring the inversion of a smaller matrix.

Equation (12) gives us the entire *a posteriori* probability distribution of $\vec{a}$, *i.e.*, the conditional probability density of $\vec{a}$ given the data and our assumed power spectrum. This probability density is a multivariate Gaussian with mean $\vec{a}_{\mathrm{ML}}$ and covariance matrix

$$\mathbf{K} = \left(\mathbf{C}^{-1} + \mathbf{W} \cdot \mathbf{Y}^T \cdot \mathbf{N}^{-1} \cdot \mathbf{Y} \cdot \mathbf{W}\right)^{-1}. \tag{15}$$

This probability distribution tells us how likely a particular value of $\vec{a}$ is, given our assumptions. The square roots of the diagonal elements of this matrix can be interpreted as standard error estimates for the elements of $\vec{a}_{\mathrm{ML}}$, and those of $\mathbf{Y} \cdot \mathbf{K} \cdot \mathbf{Y}^T$ as error estimates for the elements of $\vec{s}_{\mathrm{ML}} = \mathbf{Y} \cdot \vec{a}_{\mathrm{ML}}$, which is the maximum-likelihood value of the true CMB signal. In addition, armed with the full covariance matrix $\mathbf{K}$, we can construct constrained realizations of the microwave sky by simply choosing Gaussian random variables from the appropriate distribution. In this way we can construct simulated sky maps with a probability distribution that correctly incorporates both our presumed model and the existing data. These simulated maps can be used to make predictions for other experiments, as we shall see below.

Unfortunately, equation (3) is not a completely adequate model of the DMR data, because the data do not contain information about all of the multipole moments. In particular, the data are completely insensitive to the monopole contribution, and the dipole is contaminated by the much larger kinematic dipole due to our motion with respect to the CMB rest frame. The Wiener filter derived above must be modified to account for this. For simplicity of notation, we will ignore beam-smoothing by setting $\mathbf{W} = \mathbf{1}$ throughout the remainder of this section.

Suppose that the data contain no reliable information about modes with $l \leq l_*$. To account for the monopole and dipole, we should set $l_* = 1$; if we decided to discard the quadrupole as well, we would set $l_* = 2$. Then we should modify equation (3) as follows:

$$\vec{d} = \mathbf{Y} \cdot \vec{a} + \mathbf{Z} \cdot \vec{b} + \vec{n}, \tag{16}$$

where we have excised the elements of $\vec{a}$ and the columns of $\mathbf{Y}$ corresponding to the modes about which we have no information. The dimension of $\vec{a}$ is now $\mathcal{M} = (l_{\max} + 1)^2 - (l_* + 1)^2$ instead $(l_{\max} + 1)^2$, and the corresponding dimensions of $\mathbf{Y}$ and $\mathbf{C}$ are reduced in the same way. $\mathbf{Z}$ is an $\mathcal{N} \times (l_* + 1)^2$ matrix containing the columns removed from $\mathbf{Y}$, and $\vec{b}$ is an $(l_* + 1)^2$-dimensional vector containing the unknown coefficients of the excluded modes.

We can still try to choose our filter $\mathbf{F}$ to be a least-squares estimator in the sense of equation (9). However, we must impose an additional constraint on $\mathbf{F}$: since we have no knowledge at all of the true value of $\vec{b}$, we should demand that the result $\mathbf{F} \cdot \vec{d}$ of filtering should be independent of $\vec{b}$. We can do this by imposing the constraint

$$\mathbf{F} \cdot \mathbf{Z} = 0. \tag{17}$$

If we solve this constrained minimization problem by the usual method of Lagrange multipliers, we find that

$$\mathbf{F} = \mathbf{Y} \cdot \mathbf{C} \cdot \mathbf{Y}^T \cdot \mathbf{M}^{-1}(\mathbf{1} - \mathbf{Z} \cdot (\mathbf{Z}^T \cdot \mathbf{M}^{-1} \cdot \mathbf{Z})^{-1} \cdot \mathbf{Z}^T \cdot \mathbf{M}^{-1}) \equiv \mathbf{Y} \cdot \mathbf{C} \cdot \mathbf{Y}^T \cdot \mathbf{M}^{-1} \cdot \mathbf{P}. \quad (18)$$

This equation differs from the original filter (10) only in the projection operator $\mathbf{P}$. This operator simply projects the data vector onto a subspace orthogonal to the unwanted modes, where orthogonality is determined with respect to the inner product $\langle \vec{x}, \vec{y} \rangle = \vec{x}^T \cdot \mathbf{M}^{-1} \cdot \vec{y}$.

The Wiener filter (18) is also a maximum-likelihood filter: the conditional probability distribution for $\vec{a}$ and $\vec{b}$ given the data is $P(\vec{a}, \vec{b} \,|\, \vec{d}) \propto \exp\left(-\frac{1}{2}\chi^2\right)$, with

$$\begin{aligned}\chi^2 &= (\vec{a} - \vec{a}_{\mathrm{ML}})^T \cdot (\mathbf{C}^{-1} + \mathbf{Y}^T \cdot \mathbf{N}^{-1} \cdot \mathbf{Y}) \cdot (\vec{a} - \vec{a}_{\mathrm{ML}}) \\ &+ (\vec{b} - \vec{b}_{\mathrm{ML}})^T \cdot \mathbf{Z}^T \cdot \mathbf{M}^{-1} \cdot \mathbf{Z} \cdot (\vec{b} - \vec{b}_{\mathrm{ML}}) + \vec{d}^T \cdot \mathbf{M}^{-1} \cdot \vec{d},\end{aligned} \quad (19)$$

where

$$\vec{a}_{\mathrm{ML}} = \mathbf{C} \cdot \mathbf{Y}^T \cdot \mathbf{M}^{-1} \cdot \mathbf{P} \cdot \vec{d}, \quad (20)$$
$$\vec{b}_{\mathrm{ML}} = (\mathbf{1} - \mathbf{P}) \cdot \vec{d}. \quad (21)$$

The maximum-likelihood filter $\mathbf{F}'$ in equation 14 must therefore be modified by including the same projection operator $\mathbf{P}$, and the least-squares and maximum-likelihood filters still coincide. Furthermore, equation (15) for the covariance matrix $\mathbf{K}$ in the *a posteriori* probability density of $\vec{a}$ is unchanged.

We conclude this section with a pragmatic note. It is in fact not generally necessary to compute the projection operator $\mathbf{P}$ in order to properly project out the unwanted modes. Rather, one can achieve identical results by simply including the unwanted modes together with the others, so that $\vec{a}$ is once again an $(l_{\max}+1)^2$-dimensional vector. We extend the signal covariance matrix $\mathbf{C}$ by setting $C_{\mu\nu} = \infty \delta_{\mu\nu}$ in the rows and columns corresponding to the unwanted modes. Then the filter $\mathbf{F}$ and the covariance matrix $\mathbf{K}$ computed in the previous section give precisely the right results.[1]

This result is easy to justify heuristically: since we have no information about the coefficients $\vec{b}$, it is natural to assign them an infinitely wide prior distribution. It is also straightforward to verify by a direct computation that the results of the previous section approach the results of this section as the unknown elements of $\mathbf{C}$ tend to infinity (Rybicki & Press 1992).

## 3 Results

In this section we will apply the filter derived above to the two-year DMR data. Figure 1 shows the result of applying the Wiener filter to the two-year DMR data. The data set used to make this map is a weighted average of the 53 and 90 GHz ecliptic-projected two-year sky maps. The weights were chosen to be inversely proportional to the noise variances in the maps in order to minimize the noise in the average. The usual 20° Galactic cut was imposed on the map before filtering. The input power spectrum was a standard cold dark matter power spectrum with normalization $Q = 20 \, \mu\text{K}$. Varying the input power spectrum leads to slight variations in the amount of small-scale power in the filtered map, but the large-scale features are essentially unchanged. It comes as no surprise that there is less information in the Galactic plane than at high Galactic latitude.

---

[1] One does not of course set anything equal to infinity in one's code: matrices with infinite entries are notoriously ill-conditioned. In practice, if we use equation (14) to compute $\mathbf{F}$, we need only $\mathbf{C}^{-1}$, never $\mathbf{C}$, and we can simply set the appropriate entries of $\mathbf{C}^{-1}$ to zero.

The root-mean-square anisotropy level in the filtered map in Figure 1 is $30.6 \, \mu\text{K}$. The ensemble-average expected signal for the power spectrum we have chosen is $42.5 \, \mu\text{K}$, so we have reason to believe that we have recovered a significant fraction of the power. We can assess the uncertainty in this filtered map in a variety of ways. Given the data and the assumed power spectrum, the *a posteriori* probability distribution of the true signal is Gaussian with mean equal to the filtered signal. The simplest way to assign uncertainties is simply to compute the standard deviation of this probability distribution. In other words, we compute the square roots of the diagonal elements of the covariance matrix $\mathbf{Y} \cdot \mathbf{K} \cdot \mathbf{Y}^T$. The ratio of the signal to the uncertainty is shown in Figure 2.

We can get a qualitative idea of the statistical significance of features in the map from this figure: individual pixels in the most prominent hot and cold spots are significant at the three to four sigma level. Of course, the various pixels are not independent: the covariance matrix $\mathbf{Y} \cdot \mathbf{K} \cdot \mathbf{Y}^T$ has off-diagonal elements. The correlation between the errors in nearby pixels is approximately Gaussian with standard deviation 5°.7. We can also get a handle on the degree of uncertainty in the filtered maps by making constrained realizations of the temperature anisotropy in the manner described in the previous section. Four such realizations are shown in Figure 3.

We can define the angular power spectrum of the filtered map:

$$A_l = \frac{1}{2l+1} \sum_{m=-l}^{l} (a_{\text{ML}})_{lm}^2. \tag{22}$$

This power spectrum, shown in Figure 4, s a useful indicator of the amount of information recovered by the filtering process. On large scales the recovered power $A_l$ is comparable to the input power spectrum, while for modes with high $l$ the power is nearly zero. The Wiener filter, by construction, returns zero values for modes for which we have no information: after all, in the absence of any information from the data, the *a posteriori* probability distribution for $a_{lm}$ is the same as the prior distribution, which is a Gaussian of zero mean.

## 4   Predictions Based on the Wiener-Filtered Map

One use for the Wiener-filtered DMR maps is to make predictions for other experiments. We can process the filtered map through the window function of the other experiment to determine the most probable signal for the experiment to observe, given the DMR data and the assumed power spectrum. Furthermore, since we know the entire conditional probability distribution for $\Delta T$ given the DMR data and the assumed prior distribution, we can compute the probability distribution for the predicted signal. Since we have incorporated the DMR information, this distribution will in general be narrower than the *a priori* probability distribution associated with our theoretical model. We should therefore be able to place tighter constraints on our model than we could without using the DMR data. By using the filtered data rather than the raw data, we reduce the uncertainty in the prediction as well as avoiding the technical difficulties associated with comparing data sets of different resolutions and beam-switching strategies (Lineweaver et al. 1995). If one has access to both data sets, of course, then both can be used as inputs to the Wiener filter, rather than using the data from one experiment to make predictions for the other.

Of course, we only expect to gain significantly if we choose an experiment whose window function has significant overlap with the DMR window function, so that the modes that we have estimated well from the DMR data make a substantial contribution to the signal in the other experiment. The Tenerife experiment would seem to be ideal for this purpose, since its beam size of $5°.2$ (FWHM) is only slightly smaller than the $7°$ DMR beam. Unfortunately, the fact that Tenerife is a triple-beam experiment works against us, since the sensitivity to modes with low $|m|$ is quadratically suppressed. This fact significantly reduces the amount of information we can recover from the filtered DMR map.

In Figure 5 we show the COBE and Tenerife window functions $\mathcal{W}_l$. $\mathcal{W}_l$ is the window function used to compute the mean-square signal expected in an experiment. Specifically, if $R(\hat{\mathbf{r}})$ is the response of the experiment when pointed at a particular position $\hat{\mathbf{r}}$, then

$$\langle R^2 \rangle = \frac{1}{4\pi} \sum_l (2l+1) C_l \mathcal{W}_l. \tag{23}$$

Note that this is different from the quantity $W_l$ that we defined above: $W_l$ is used to compute the response $R$, not the mean-square response $\langle R^2 \rangle$. For COBE, the two are related in a simple manner: $\mathcal{W}_l = W_l^2$. For Tenerife, the beam-chopping breaks the circular symmetry, and so $W$ depends on both $l$ and $m$ (see below). In this case,

$$\mathcal{W}_l = \frac{4\pi}{2l+1} \sum_{m=-l}^{l} W_{lm}^2 |Y_{lm}(\hat{\mathbf{r}})|^2. \tag{24}$$

The Tenerife experiment has taken data covering the range $140°$ to $260°$ in right ascension and $35°$ to $45°$ in declination. Qualitative maps of the entire data set have been published (Rebolo et al. 1995), but only the strip near $40°$ has been quantitatively analyzed (Watson et al. 1992). We will attempt to predict as much as we can of the Tenerife signal over the entire range. The strip of sky observed by the experiment is shown in Figure 1.

The process of making predictions is fairly simple. We take the spherical harmonic coefficient $\vec{a}_{\text{ML}}$ and multiply it by the Tenerife window function $W^{(\text{Ten})}$:

$$W_{lm}^{(\text{Ten})} = (1 - \cos m\alpha) \exp(-\tfrac{1}{2}\sigma_{\text{Ten}}^2 l(l+1)). \tag{25}$$

Here $\sigma_{\text{Ten}} = 2°.2$ is the Tenerife beam size, and $\alpha \sin\theta = 8°.1$ is the amplitude of the beam chop.[2] The resulting vector contains the coefficients of a spherical harmonic expansion of the most

---

[2] The window function is only simple in a coordinate system in which the chop is in the azimuthal direction. We must either compute $\vec{a}_{\text{ML}}$ in these coordinates in the first place or else rotate to them afterwards. The Tenerife experiment always chops in right ascension, so the rotation is easy to perform.

probable *a posteriori* Tenerife signal $R_{\rm ML}$:

$$R_{\rm ML}(\theta, \varphi) = \sum_{l,m} W^{({\rm Ten})}_{lm} (a_{\rm ML})_{lm} Y_{lm}(\theta, \varphi). \qquad (26)$$

The result is shown in Figure 6. The portion of sky plotted corresponds to the region of sky covered by the Tenerife experiment. The root-mean-square value of $R_{\rm ML}$ over this region is 16.3 $\mu$K. For the standard CDM model we are considering, the expected r.m.s. signal is 32.7 $\mu$K, and so it appears that we have predicted a significant fraction of the total signal. Recall that the Wiener filter returns low values for modes for which it has little information, and values near the expected value for modes with high signal-to-noise ratio. The ratio of the Wiener-filtered prediction to the *a priori* expectation value is thus a rough indicator of how much signal has been recovered.

We can of course do much better than this qualitative argument. We know that the vector $\vec{a}$ is Gaussian with mean $\vec{a}_{\rm ML}$ and covariances given by (15). The Tenerife response $R$ is therefore Gaussian distributed with mean $R_{\rm ML}$ and covariance

$$\text{Cov}(R(\hat{\mathbf{r}}_1), R(\hat{\mathbf{r}}_2)) = \sum_{\mu,\nu} W^{({\rm Ten})}_\mu Y_\mu(\hat{\mathbf{r}}_1) W^{({\rm Ten})}_\nu Y_\nu(\hat{\mathbf{r}}_2) K_{\mu\nu}. \qquad (27)$$

The standard error associated with each point in Figure 6 is therefore $\sigma(\hat{\mathbf{r}}) = \sqrt{\text{Cov}(R(\hat{\mathbf{r}}), R(\hat{\mathbf{r}}))}$. The standard error ranges from 26.4 $\mu$K to 29.7 $\mu$K.

One way to get an idea of the amount of information we have gained by this exercise is to compare these standard errors with the level of uncertainty we would have without the DMR data. If we made no use of the DMR data, then our probability distribution for $R(\hat{\mathbf{r}})$ would simply be Gaussian with zero mean and standard error given by the r.m.s. expected signal of 32.7 $\mu$K. Using the DMR data reduces the standard error by about 15%. This amounts to at least a modest improvement in our ability to assess the consistency of this model with the Tenerife data. Furthermore, some particular features in the prediction, particularly the two positive features at large right ascension, have quite high significance.

In Figure 7 we have plotted the predicted signal and associated errors in several individual strips for greater ease of comparison with the actual data. The maximum at right ascension of 250° coincides almost exactly with the largest peak in the Tenerife map (Rebolo et al. 1995).

The standard errors simply represent the diagonal part of the covariances in equation (27). For a complete statistical description of the prediction, we need to know the correlations between distinct points as well. Because of the beam-switching strategy, these correlations can be quite important. The general shape of the correlations is as one would expect from the beam-switching patterns: a positive lobe flanked by two symmetric negative lobes. The following fitting formula reproduces the three main lobes to about 5% accuracy, although it misses small ripples at high separation.

$$\text{Cov}(R(\hat{\mathbf{r}}_1)R(\hat{\mathbf{r}}_2)) = \text{Cov}(R(\hat{\mathbf{r}}_1)R(\hat{\mathbf{r}}_1))e^{-\Delta\delta^2/2(4°\!.7)^2} \times$$
$$\left(e^{-\Delta A^2/2(3°\!.0)^2} - 0.69 e^{-(\Delta A - 9°\!.2)^2/2(3°\!.0)^2} - 0.69 e^{-(\Delta A + 9°\!.2)^2/2(3°\!.0)^2}\right), \qquad (28)$$

where $\Delta\delta$ is the declination difference between $\hat{\mathbf{r}}_1$ and $\hat{\mathbf{r}}_2$, and $\Delta A = \Delta\varphi \cos \delta_1$ is the separation in azimuth.

As with the original Wiener-filtered map, it may be helpful to look at constrained realizations of the statistical fluctuations about the maximum-likelihood Tenerife prediction in order to get some idea of the level of uncertainty. Five such realizations are plotted in Figure 8. We show both greyscale maps for comparison with Figure 6 and linear plots of the strip at declination 35° for comparison with Figure 7.

The Wiener filter depends on the input power spectrum assumed. It is therefore of interest to ask how robust the features in the Tenerife predictions are under changes in this input power spectrum. We have repeated our analysis using input power spectra with primordial power spectra having slopes $n = 1.5$ and $n = 0.5$ instead of the standard value $n = 1$. To give an idea of the magnitude of this variation in $n$, we remind the reader that the COBE data determine $n$ with a one-sigma uncertainty of about 0.3 (Tegmark & Bunn 1995). The amplitude of the predicted structure increases with increasing $n$, as one would expect. When we vary $n$ by $\pm 0.5$, the two highest peaks in the lowest panel of Figure 7 change by $\pm 11\,\mu$K and $\pm 9\,\mu$K, or approximately 15-20%. The location of the peaks does not change significantly. The presence and approximate height of these two peaks can therefore be regarded as robust predictions.

## 5 Conclusions

Wiener filtering is a promising tool for the analysis of CMB sky maps. Filtering provides a significant improvement in the signal-to-noise ratio in the regions covered by the raw data, and allows some small amount of information to be reconstructed about the anisotropy within the Galactic cut region. As Figure 2 shows, the most prominent hot and cold spots in the filtered map carry very high statistical significance.

This reduction in noise is not without a price. In order to apply the Wiener filter, one needs to assume a power spectrum, and the cleaned data therefore depend on more assumptions than do the raw data. However, the large-scale features in the filtered map are only weakly dependent on the choice of power spectrum.

One great advantage of the Wiener-filtered map is that the statistical properties of the residuals are known precisely, assuming the underlying model is correct. It is therefore possible to assess the goodness of fit of any other data set with the combined DMR data and assumed theory. In addition, it is easy to make constrained realizations of the microwave sky with the correct *a posteriori* probability distribution. This technique can in principle be used to perform Monte-Carlo simulations of the CMB for comparison with experiments on similar angular scales.

The Tenerife experiment seems to be the most promising experiment on which to test the possibility of using the Wiener filtered maps to make predictions. Unfortunately, the triple-beam strategy used in the experiment drastically reduces Tenerife's sensitivity to the modes that are constrained well by the DMR. It is possible, however, to predict some fraction of the Tenerife signal from the filtered DMR data. By incorporating this information into a Tenerife analysis, it should be possible to place somewhat tighter constraints on models than would be possible with the Tenerife data alone.

In particular, the DMR data predict two strong features in the Tenerife data at right ascensions around 220° and 255° and declinations around 35° (see Figures 7 and 8). The statistical significance of these features is high enough that if they are not found to be present in the real data, we will have strong reason to conclude that something is wrong with either the assumptions underlying the prediction or the data from one of the experiments. Specifically, if the Tenerife experiment were to fail to see these features, we would be led to one of the following conclusions:

1. The CMB anisotropy is not well described by Gaussian statistics. (The Wiener filtered map does not depend on the assumption of Gaussian statistics, but the uncertainties in it do.)

2. The CMB anisotropy has a power spectrum drastically different from that usually assumed.

3. One of the two experiments is contaminated by systematic error or foregrounds.

The predicted features are fairly robust to reasonable changes in the CMB power spectrum on COBE scales, so the only way the second option could be correct is if the power spectrum has a drastic rise at high enough $l$ to be invisible to COBE. It is somewhat difficult to see how the third option could be correct: the COBE data have been extensively checked for systematic errors and foregrounds (Bennett et al. 1992, Kogut et al. 1992), and it would be difficult to explain the failure of Tenerife to see the predicted signal by assuming a contamination of the Tenerife data: contaminants are much more likely to induce spurious signals than to mask real ones. Naturally, any of these three conclusions would be of enormous interest.

We would like to thank M. White for supplying us with the CDM angular power spectrum. Y.H. acknowledges the hospitality of the Center for Particle Astrophysics. Y.H. has been supported in part by The Hebrew University Internal Funds (grant 53/94) and by the Israel Science Foundation administered by the Israeli Academy of Sciences and Humanities (grant 590/94).

**Figure Captions**

Figure 1: The Wiener-filtered DMR data

The result of applying a Wiener filter to the two-year COBE DMR data. The input power spectrum is that of a standard cold dark matter model. The contour levels are at $\pm 25, \pm 50, \pm 75\,\mu$K, with negative values indicated by dashed lines. The region of sky observed by the Tenerife experiment, *i.e.*, the region with equatorial coordinates $140° <$ R.A. $< 260°$ and $35° < \delta < 45°$, is also marked on the greyscale map. In Section 4 we will use the filtered DMR map to make predictions for this experiment. The part of the strip near $260°$ in right ascension is split in half in this projection.

Figure 2: Signal-to-noise ratio in the filtered map

The ratio of the filtered map in Figure 1 to the uncertainty is plotted. The contours take values $\pm 2, \pm 3, \pm 4$, with dashed lines representing negative values.

Figure 3: Constrained realizations

Four constrained realizations of the microwave sky are shown. The realizations represent typical expected fluctuations about the Wiener-filtered map in Figure 1, and were drawn from the Gaussian probability distribution described in the previous section. Each realization is shown in both greyscale and contour maps. The contour levels are $\pm 50, \pm 100, \pm 150\,\mu$K, with negative values indicated by dashed lines.

Figure 4: Power spectrum of the Wiener-filtered map

The power spectrum $A_l$ defined in equation (22) is shown. The standard CDM input power spectrum is also plotted for comparison.

Figure 5: DMR and Tenerife window functions.

The power-spectrum window functions $\mathcal{W}_l$ are shown for both the COBE and Tenerife experiments. The one with little power at low $l$ is Tenerife. The dashed line shows $l(l+1)C_l$, where $C_l$ is an arbitrarily-normalized standard cold dark matter power spectrum.

Figure 6: Tenerife predictions

A map of the most probable (*a posteriori*) Tenerife signal over the region $120° <$ RA $< 260°$, $35° < \delta < 45°$. The greyscale ranges from $-60\,\mu$K to $60\,\mu$K, and the uncertainty associated with each point is approximately $28\,\mu$K.

Figure 7: Tenerife predictions with uncertainties

The solid lines show slices through the map in Figure 6, and the dashed lines show plus and minus one-sigma errors. The top, middle,and bottom panels correspond to declinations of 45°, 40°, and 35° respectively.

Figure 8: Constrained realizations of the Tenerife data

We show five constrained realizations of the Tenerife predictions to assist the reader in qualitatively assessing the uncertainties in the predictions in Figures 6 and 7. The top panel shows five realizations covering the same range as Figure 6. The greyscale in this map ranges from -110 to 100 $\mu$K. The bottom panel is a strip through the five realizations at declination 35°, and is thus to be compared with the lowest panel of Figure 7.